\makeatletter
\newcommand*{\addFileDependency}[1]{% argument=file name and extension
	\typeout{(#1)}
	\@addtofilelist{#1}
	\IfFileExists{#1}{}{\typeout{No file #1.}}
}
\makeatother

\newcommand*{\myexternaldocument}[1]{%
	\externaldocument{#1}%
	\addFileDependency{#1.tex}%
	\addFileDependency{#1.aux}%
}

\documentclass[aps,amsmath,amssymb,nofootinbib,showpacs,superscriptaddress]{revtex4-2}
\usepackage[english]{babel}
\usepackage{latexsym}
\usepackage{graphics}
\usepackage{graphicx}
\usepackage{epsfig}
\usepackage{color}
\usepackage{bm}
\usepackage{amsmath}
\usepackage{amssymb}
\usepackage{amsthm}
\usepackage{dcolumn}
\usepackage{bm}
\usepackage{float}
\usepackage{xr-hyper}
\usepackage{color}
\usepackage{epstopdf}
\usepackage{cleveref}
\usepackage[autostyle]{csquotes}
\usepackage{soul,color}
\usepackage[svgnames]{xcolor}

\myexternaldocument{Supporting}

\theoremstyle{remark}

\begin{document}
	
	\renewcommand{\thefootnote}{\fnsymbol{footnote}}
	
	\preprint{APS/123-QED}
	
	\title{Machine-learning Driven Synthesis of TiZrNbHfTaC5 High-Entropy Carbide}
	%\title{Targeted Vacuumless Synthesis of Single-Phase High-Entropy Carbide TiZrNbHfTaC$_{5}$ based on Computational Prediction}
	
	\author{Alexander Ya. Pak} 
	\affiliation{National Research Tomsk Polytechnic University, 30 Lenin Avenue, Tomsk 634050, Russia}
	
	\author{Vadim Sotskov\^* }
	\affiliation{Skolkovo Institute of Science and Technology, Skolkovo Innovation Center, Bolshoi Blv. 30, Building 1, Moscow 121205, Russia}
	
	\author{Arina A. Gumovskaya} 
	\affiliation{National Research Tomsk Polytechnic University, 30 Lenin Avenue, Tomsk 634050, Russia}
	
	\author{Yuliya Z. Vassilyeva} 
	\affiliation{National Research Tomsk Polytechnic University, 30 Lenin Avenue, Tomsk 634050, Russia}
	
	\author{Zhanar S. Bolatova} 
	\affiliation{National Research Tomsk Polytechnic University, 30 Lenin Avenue, Tomsk 634050, Russia}
	
	\author{Yulia A. Kvashnina} 
	\affiliation{Pirogov Russian National Research Medical University, 1 Ostrovityanova St., Moscow 117997, Russia}
	
	\author{Gennady Ya. Mamontov} 
	\affiliation{National Research Tomsk Polytechnic University, 30 Lenin Avenue, Tomsk 634050, Russia}
	
	\author{Alexander V. Shapeev}
	\affiliation{Skolkovo Institute of Science and Technology, Skolkovo Innovation Center, Bolshoi Blv. 30, Building 1, Moscow 121205, Russia} 
	
	\author{Alexander G. Kvashnin\^*, }
	\affiliation{Skolkovo Institute of Science and Technology, Skolkovo Innovation Center, Bolshoi Blv. 30, Building 1, Moscow 121205, Russia}
	
	\footnotetext{Vadim.Sotskov@skoltech.ru, \\  A.Kvashnin@skoltech.ru}
	
	\date{\today}
	\begin{abstract}
		Synthesis of high-entropy carbides (HEC) requires high temperatures that can be provided by electric arc plasma method.
		However, the formation temperature of a single-phase sample remains unknown. Moreover, under some temperatures multi-phase structures can emerge.
		In this work we developed an approach for a controllable synthesis of HEC TiZrNbHfTaC5 based on theoretical and experimental techniques.
		We used canonical Monte Carlo (CMC) simulations with the machine learning interatomic potentials to determine the temperature conditions for the formation of single-phase and multi-phase samples.
		In full agreement with the theory, the single-phase sample, produced with electric arc discharge, was observed at 2000 K. Below 1200 K the sample decomposed into (Ti-Nb-Ta)C and a mixture of (Zr-Hf-Ta)C, (Zr-Nb-Hf)C, (Zr-Nb)C, and (Zr-Ta)C. Our results demonstrate the conditions for the formation of HEC and we anticipate that our approach can pave the way towards targeted synthesis of multicomponent materials. 
	\end{abstract}
	
	\maketitle
	
	\section{Introduction} 
	Ultra-High Temperature Ceramics (UHTC) are a class of refractory ceramics that are stable at high temperature, and they are usually made of carbides, borides, or nitrides of IV and V group transition metals.
	Over the past few years, both theoretical and experimental development of refractory high-entropy materials has been actively pursued, including high-entropy carbides (HEC)\cite{castle_processing_2018, oses_high-entropy_2020, hossain_carbon_2021, hossain_entropy_2021, potschke_preparation_2021}.
	HECs are equimolar multi-component single-phase solid solution of 4--6 transition metals of IV and V group occupying cubic NaCl-type crystalline lattice.
	
	One of the main factors responsible for stabilization of HECs is configurational (or mixing) entropy, which should be more than $1.5R$, i.e. at least five principal elements in the composition \cite{yeh_nanostructured_2004,miracle_critical_2017,george_high-entropy_2019}.
	Theoretical studies, including the use of machine learning methods, allowed the prediction of existence of several dozen compositions of single crystal phases of high-entropy carbides \cite{sarker_high-entropy_2018, harrington_phase_2019, kaufmann_discovery_2020}.
	Despite of the configurational entropy there are a number of compositions with a sufficient number of elements that do not form a single-phase high-entropy material \cite{sarker_high-entropy_2018}.
	It was theoretically shown that mixing enthalpy, electronegativity, and valence electron configuration can help to explain the single-phase nature of an high-entropy materials \cite{oh_engineering_2019,tang_role_2022,ji_relative_2015,liu_phase_2021,guo_effect_2011}.
	
	The most common method to synthesize high-entropy carbides is reactive spark plasma sintering (SPS) of pre-homogenized raw materials based on individual metal carbides, pure metals or metal oxides \cite{demirskyi_synthesis_2020,wei_high_2019, li_phase_2021}.
	In some cases, liquid precursors can be used to ensure homogeneity of a raw material \cite{li_liquid_2019}.
	Usually, the synthesis of HEC is performed at high temperatures about 2200--2300$^{\circ}$C, and SPS is usually realized at pressures of about 10--60 MPa with a 10--15 min dwell.
	At the same time the homogenization of a raw material can be made for more than one day \cite{dusza_microstructure_2018,wang_irradiation_2020,han_improved_2020}. 
	
	Despite of special conditions, the synthesis of a single-phase HEC cannot be performed succesfully every time and the conditions which are responsible for the synthesis of single- or multi-phase samples have not been studied yet.
	At the same time there has been no attempts towards the synthesis of all HECs predicted by various theoretical techniques.
	To quickly test hypotheses about the possibility to synthesize a particular material a simple, cheap, and efficient method providing high temperatures for successful synthesis is required.
	
	Electric arc plasma methods look promising in the synthesis of high-entropy materials primarily because of the possibility of reaching high temperatures, and ensuring high heating rates \cite{zhang_arc_2019,wei_high-temperature_2021,yao_carbothermal_2018}. 
	These methods have been already employed for the synthesis of transition metal carbides \cite{saito_encapsulation_1993,saito_encapsulation_1997}. 
	In the last few years, the so-called vacuumless electric arc synthesis method has been actively developed, which involves initiation of a direct current arc discharge between graphite electrodes located in an autonomous gas environment.
	During this process, the synthesis is realized in an autonomous gas environment containing CO and CO$_{2}$, which prevents oxidation of products of reaction \cite{pak_novel_2022,pak_vacuumless_2020}.
	This approach allows one to simplify experimental scheme, reduce energy intensity of the synthesis process, increase efficiency as shown by the synthesis of carbon nanostructures of some carbides \cite{li_synthesis_2010,a_synthesis_2015,zhao_continuous_2012,su_low-cost_2014,pak_synthesis_2022}. 
	
	Electric arc methods can be useful in the development of approaches to synthesize high-entropy carbides because of the wide range of achieved temperatures.
	However, the temperature conditions responsible for the formation of single- or multi-phase samples are still the subject to debate.
	Additionally, multi-component metal carbides are produced under the multiple actions of electric arc plasma \cite{zhang_arc_2019,kan_precipitation_2020,pak_synthesis_2022}, that substantially contaminate HECs by the electrode material or matrix in which the arc fusion is performed.
	Therefore, high-entropy carbides previously never dominated in the product of synthesis, obtained by the aforementioned plasma method. 
	Moreover, to our knowledge, there are no comprehensive studies of synthesis of HECs with an electric arc, that would assess the effect of other plasma treatment parameters of the raw material.
	
	To solve the issues mentioned above, we performed theoretical investigation complemented with systematic experiments for confirming the obtained theoretical data and determining the additional parameters for an effective synthesis.
	This way, we established the parameters for preparation of the initial raw material and defined the temperature regimes (according to simulation data) of single arc plasma process initiated in an autonomous gas environment to obtain single-phase HEC having the composition of TiZrNbHfTaC$_{5}$.
	
	\section{Results and Discussion} 
	\subsection{Computational determination of temperature regimes} 
	Experimental synthesis of single-phase TiZrNbHfTaC$_{5}$ HEC requires a specific temperature regime, which is usually unknown. 
	To resolve this issue we rely on our computational simulations, described in Methods section, to determine the optimal temperature range for such synthesis.  
	\\
	Before starting the simulation, we calculated total energies of the five supercells where in each of them we interchanged carbon atoms in the 4\textit{b} Wyckoff position with one of the five metallic neighbors (4\textit{a} Wyckoff position).
	After, we compared the results with the energy of the supercell where carbon atoms occupy only  4\textit{b} Wyckoff positions. 
	Eventually, we observed that the energy of the supercells with aforementioned interchanges were higher by 6 eV on average.
	Such difference reveals that energetically interchange between carbon and metallic atoms is highly unlikely and can be neglected in the CMC simulations. We therefore excluded 4$b$ carbon sites as possible sites for Monte Carlo interchange events for the CMC simulations. Thus, our CMC simulation act on the pure metallic face-centered cubic ($fcc$) sublattice of the HEC.
	Nevertheless, to account for carbon-metal interactions during CMC simulations we used the LRP trained on DFT calculations of supercells with included 4\textit{b} carbon sites.
	Further in the text we call all the simulated structures a 'high-entropy carbide', indicating the implicit presence of carbon-metal interactions in CMC simulations.
	
	\begin{figure*}
		\includegraphics[width=0.5\textwidth]{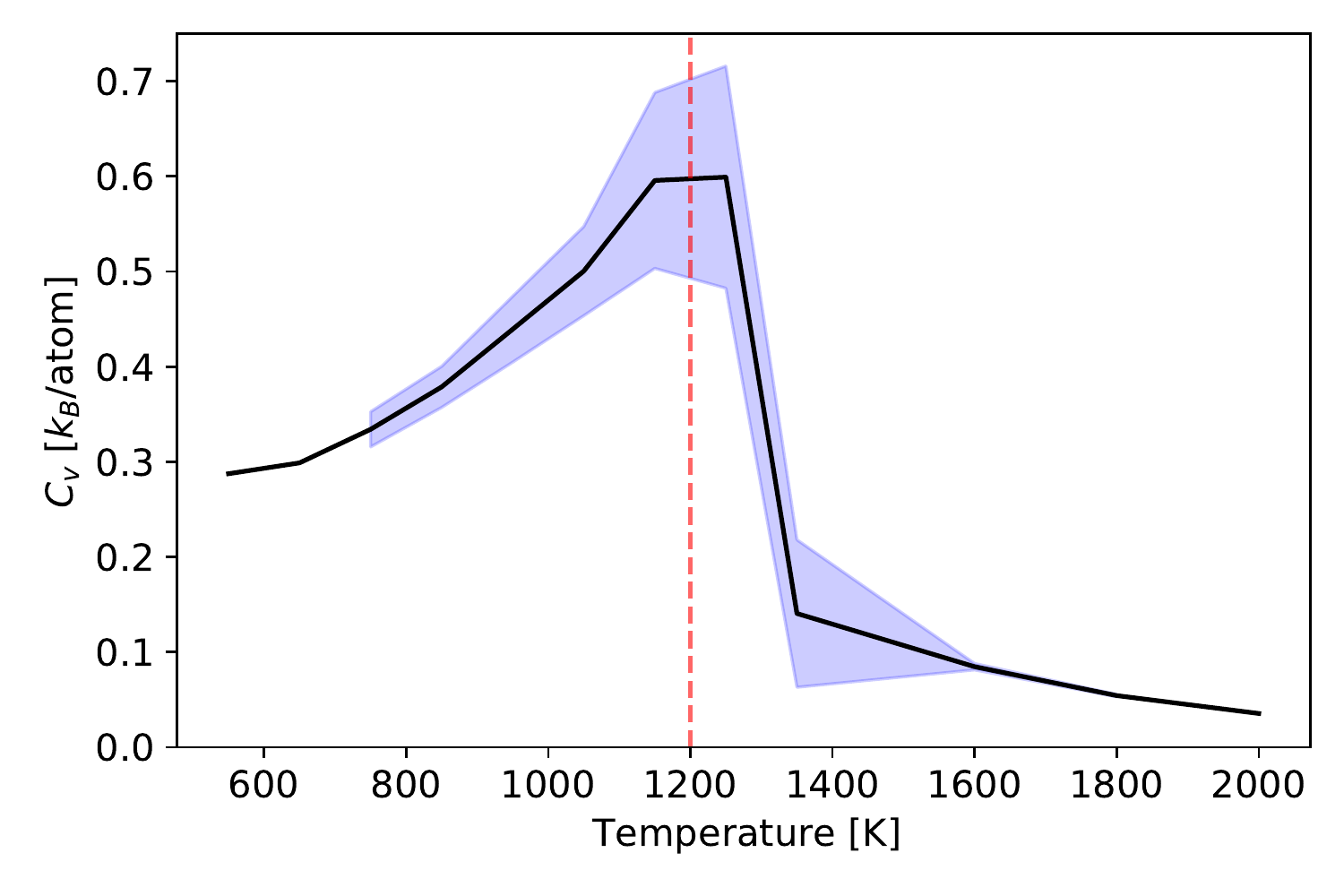}
		\caption{Specific heat capacity $C_\text{V}(T)$ for $fcc$ TiZrNbHfTaC$_{5}$ from CMC simulations for a 10 × 12 × 15 simulation box. Vertical dashed line marks the probable phase transition. Blue shaded area indicates a standard deviation calculated from 40 independent runs for each temperature}
		\label{fig:spec_heat}
	\end{figure*}
	
	The CMC simulations coupled with the LRP were performed for $10 \times 12 \times 15$ supercell (7200 metallic atoms).
	During the simulations we calculated specific heat capacity for $fcc$-TiZrNbHfTaC$_{5}$ at various temperatures in the range 500--2000 K (see Fig. 1). Detailed description of specific heat capacity calculation can be found in Methods section.
	After the simulation, we analyzed the obtained dependency of the specific heat capacity on the temperature.
	Based on this analysis, we sampled additional configurations in the vicinity of the estimated phase transition according to the workflow presented on Supplementary Fig. 1a. 
	
	Dependence of a mean value of specific heat capacity on temperature is presented by a black line on Fig. 1. Specific heat capacity was calculated between 500 K and 2000 K with a step of 150 K. After we observed a sharp decrease in $C_\text{V}$ at 1350 K, we shifted our calculations to the range between 1400 K and 2000 K with a 200 K step. The blue shaded area, that correspond to a standard deviation, indicates a higher uncertainty near a 1200 K region. Such uncertainty is determined by a significant structural differences between configurations, obtained at these temperatures. This leads to a conclusion that phase transition might be observed around 1200 K.
	We then studied the observed phase transition and possible reasons for it.
	Carefully looking at the structures simulated at temperatures above the phase transition ($T>1200$) we observed a single-phase HEC.
	On the other side, simulations at temperature regimes closer to the room temperature (T = 500 K) revealed multi-phase structure containing several multi-component carbides.
	
	To investigate the crystal structures of single- and multi-phase samples in detail, we simulated $16 \times 16 \times 16$  supercell structures (16384 metallic atoms) at $T = 500$ K and $T = 2000$ K.
	Results of simulation are presented in Fig. 2, where the crystal structures of the two simulated supercells are shown together with distributions of each constituent chemical element  among the supercell.
	Relative concentrations of chemical elements per layer along the supercell vector $b$ were calculated for both simulated structures and shown in Fig. 2c,d.
	
	\begin{figure*}
		\includegraphics[width=\textwidth]{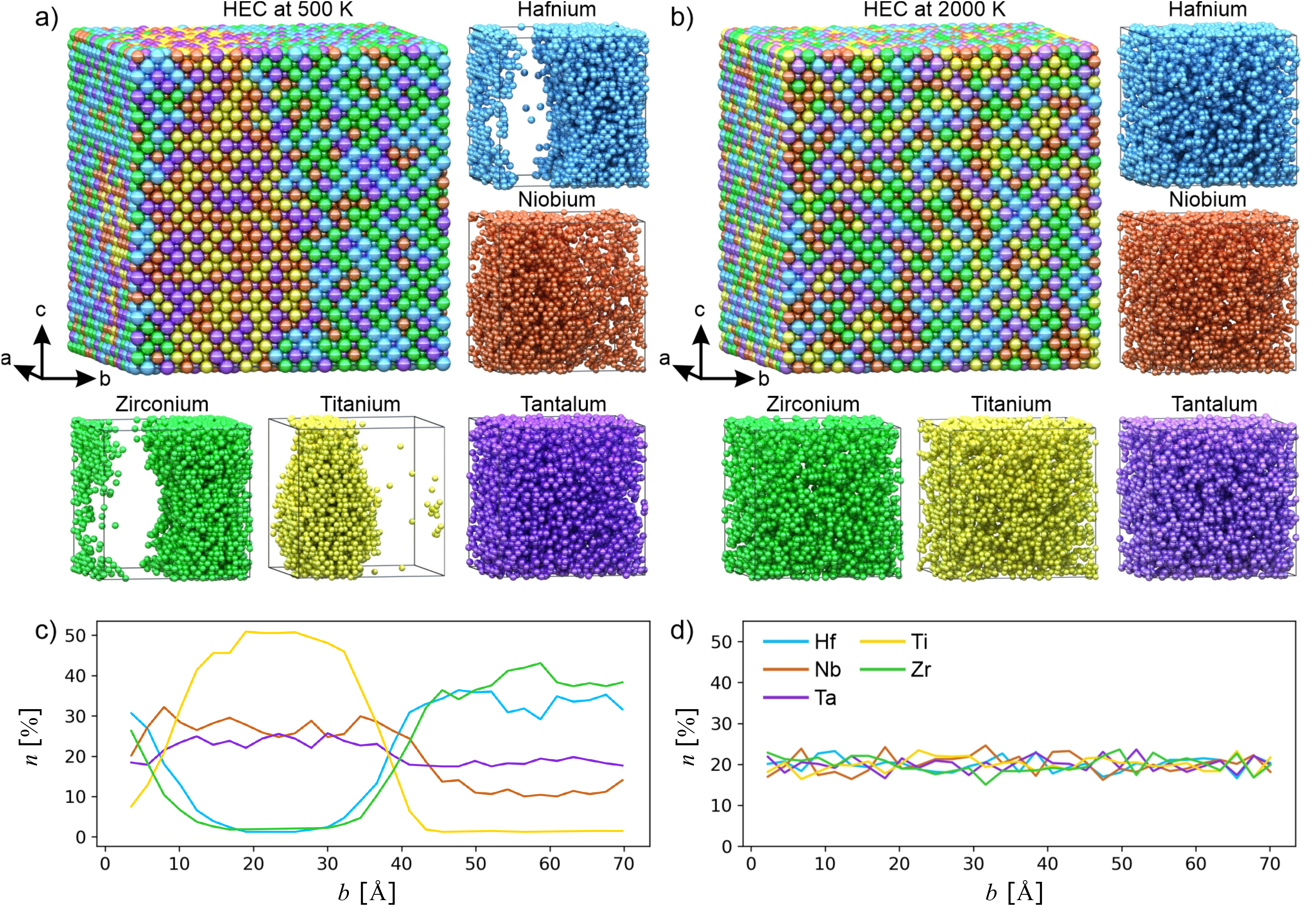}
		\caption{Crystal structures of the simulated $16 \times 16 \times 16$ supercell of a) multi- and b) single-phase (Ti-Zr-Nb-Hf-Ta)C at 500 and 2000 K respectively.
			Carbon atoms are not shown here explicitly to make the distribution of metal atoms clearer, while carbon atoms were considered in the MC simulations; c,d) relative concentration of chemical elements per layer along the supercell vector $b$}.
		\label{fig:HECs}
	\end{figure*}
	
	An interesting picture was observed for the sample at 500 K (Fig. 2a): it decomposed into separate multi-component phases.
	As it is evident from Fig. 2a,c, uniform distribution among the supercell is inherent to TaC only, while HfC and ZrC do not demonstrate a tendency to mix with TiC and NbC.
	As a result, we observed emergence of (Ti-Nb-Ta)C phase. 
	Additionally, as it is seen from Fig. 2a,c, NbC expresses a moderate ability to mix with ZrC and HfC, preferring to form a solid solution within TiC phase instead. 
	For the clarity we support this claim by a larger simulation structure of 32000 metallic atoms presented in Supplementary Fig. 2, where NbC obviously tends to accumulate in TiC regions. 
	Practically, this resulted in the emergence of the solid solution of multi-component phases phase containing Zr, Hf, Nb, and Ta.
	We preliminary assume that second phase could be a mixture of multi-component (Zr-Hf-Ta)C, (Zr-Nb-Hf)C, and (Zr-Hf-Ta-Nb)C phases with binary (Zr-Nb)C and (Zr-Ta)C.
	Decomposition into (Ti-Nb-Ta)C and other phases is shown in Fig. 2c. 
	As it is seen, the change of TaC concentration along the supercell is quite moderate and does not exceed 8\%, which is compliant with the uniform distribution of that species along the computational domain.
	On the other hand, the concentration of TiC increases in the region above 30 {\AA} and reaches maximum at 50 \AA, while concentrations of HfC and ZrC decrease towards zero within the same supercell domain. 
	The most preferable composition for second phase based on lattice parameters could be (Zr-Hf-Ta)C as the calculated lattice parameters of (Zr-Hf-Ta)C and (Ti-Nb-Ta)C are $a_{1} = 4.59$ {\AA} and $a_{2} = 4.45$ {\AA} respectively.
	More precisely this issue will be discussed below based on DFT calculations.
	
	For a single-phase HEC (Fig 2b) we predictably observed a uniform distribution of metal carbides within the supercell, that is additionally confirmed by the dispersion of concentration of metallic species, presented in Fig 2d. 
	As it is seen from this figure, deviation of atomic concentrations does not exceed 5\% on average, while the total concentration in each atomic layer amounts to around 20\% for each atomic type.
	Calculated average lattice parameter of relaxed single-phase structure is $a = 4.52$ \AA. 
	
	We suppose that the phase separation observed at 500 K in a multi-phase structure can be driven by a difference between diffusion rates of constituent metallic atoms.
	Our suggestion is supported by Ref. \cite{castle_processing_2018}, where the problem of phase separation in (Hf-Ta-Zr-Nb)C and (Hf-Ta-Zr-Ti)C mixtures was investigated. 
	Specifically, the authors studied the limiting factors for the formation of a fully mixed high-entropy carbide phase.
	They claimed that the main driving factor for the diffusion of metallic atoms is the formation of nearest-neighbor vacancy that can be occupied by diffusing atoms.
	In that case, the vacancy formation energies of host metallic atoms determine the diffusion rates and solubility of diffusing species in a host domain.
	Vacancy formation energies of five metals in the corresponding individual carbides, calculated with DFT, taken from Ref. \cite{YU201595} are presented in Supplementary Table 
	1.
	As it is seen, vacancy formation energies of Zr and Hf are almost three times higher than that of Ta. That makes TaC to act as a \enquote{host} domain for inter-diffusion of Zr and Hf atoms. 
	This leads to the formation of Zr- and Hf-based carbides.
	In a similar manner, Nb and Ti having higher vacancy formation energies can diffuse into TaC domain and form (Ti-Nb-Ta)C phase. Additionally, it was found that diffusion of metallic atoms is independent of carbon concentration \cite{YU198183, YU1979997, doi:10.1063/1.1658229, SARIAN19721637, Demaske_2017}. 
	Diffusion of carbon, in its turn, is a complicated two-step process, whose diffusion coefficients are several orders of magnitude higher than that for metals \cite{Demaske_2017}.
	Considering this, we suppose that carbon-metal interactions have insignificant effect diffusion of metal atoms.
	With the increase of temperature the leading role in phase formation is assigned to entropy processes.
	As a result of increasing entropy of mixing, a fully-mixed HEC at 2000 K is observed. 
	
	\subsection{Experimental synthesis and XRD analysis} 
	
	Results of simulations were used to perform targeted synthesis of the predicted single-phase TiZrNbHfTaC$_{5}$ HEC and a multi-phase sample.
	The synthesis was performed by using powders of pure metals of Ti, Zr, Nb, Hf, Ta. Their XRD-patterns are presented in Supplementary Fig. 3 in supporting information.
	During series of experiments with different parameters of synthesis, namely current, arc duration, milling parameters, etc., we defined the optimal ones for the formation of single-phase HEC. The detailed information about this process is presented in supporting information. The measured XRD patterns from these series of experiments are shown in Supplementary Figures 4, 5, 6, 7. They allowed us to determine the mixing parameters of the raw powders (mixing time, ball-to-powder ratio), and their electric arc treatment parameters (current and arc duration). 
	As a result we have synthesized two sets of samples containing single-phase TiZrNbHfTaC$_{5}$ and multi-phase system of two solid solutions. 
	
	\begin{figure*}
		\includegraphics[width=0.5\textwidth]{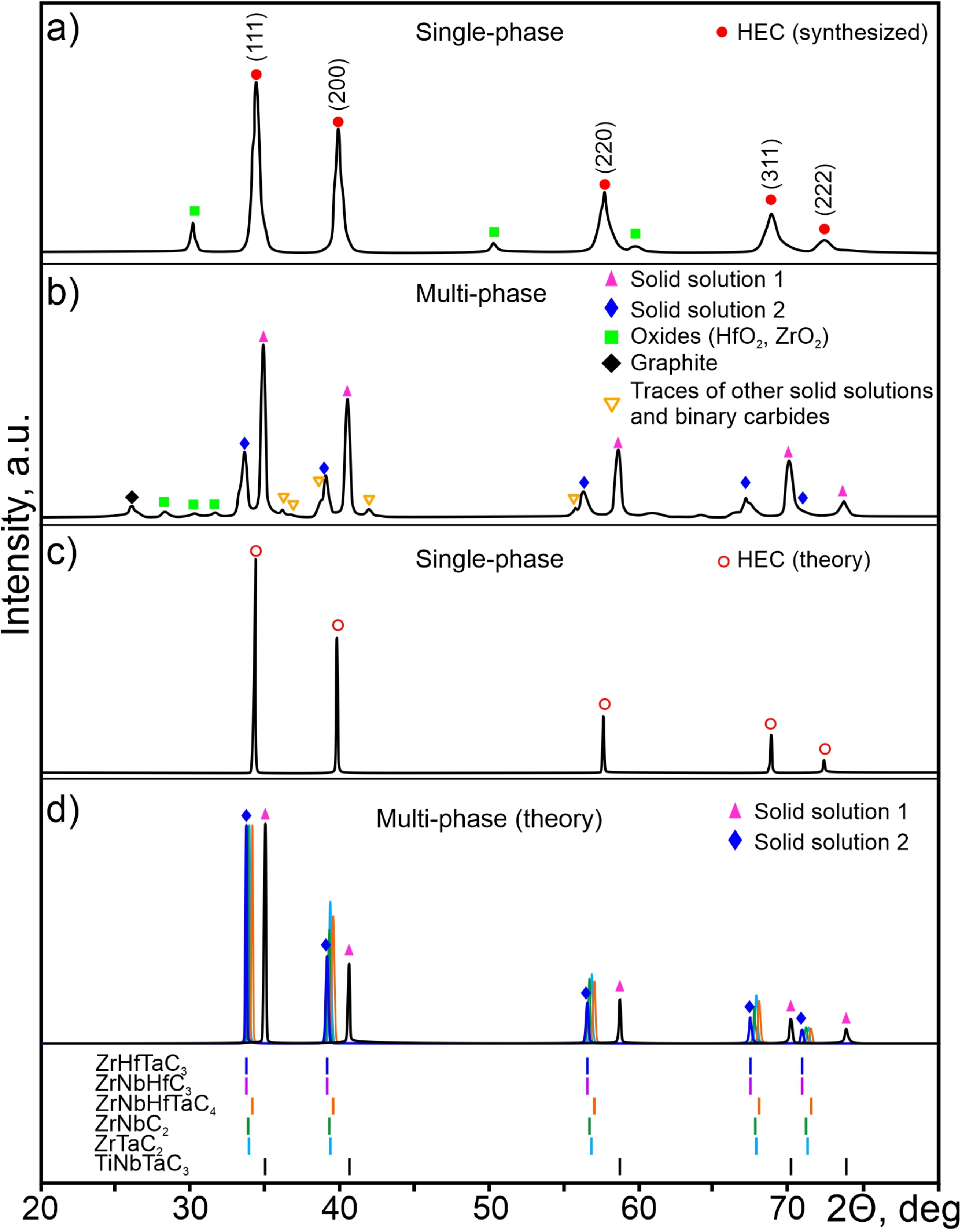}
		\caption{X-ray diffraction patterns of the synthesized a) single- and b) multi-phase samples of TiZrNbHfTaC$_{5}$, and calculated XRD patterns of c) simulated single-phase HEC, and d) separate phases of ZrHfTaC$_{3}$, ZrNbHfC$_3$, ZrNbHfTaC$_4$, ZrNbC$_2$, ZrTaC$_2$, and TiNbTaC$_{3}$ found by the analysis of calculated Gibbs free energies of mixing of various multi-component carbides.}
		\label{fig:N2}
	\end{figure*}
	
	Fig. 3 shows the measured and simulated XRD patterns of synthesized samples. As it is evident, the formation of both single- and multi-phase samples has successfully occurred.
	Moreover, we note that in comparison with the previous works in the framework of electric arc methods \cite{pak_synthesis_2022} we performed the synthesis of material with HEC as the major phase where the amount of impurities and eroded electrode material are minimized.
	Here the synthesis process is realized in a single working cycle of the electric arc reactor in an autonomous gas medium. 
	
	In the measured XRD patterns one can see symmetrical reflections corresponding to the NaCl-type crystal structure (see Fig. 3a). 
	The presence of the only phase of TiZrNbHfTaC$_{5}$ is observed.
	The lattice parameter of the synthesized single-phase TiZrNbHfTaC$_{5}$ is $a$ = 4.49 $\pm$ 0.03 \AA, which agrees well with calculated lattice parameter of the simulated HEC (4.51 \AA), which XRD pattern is shown in Fig. 3c.
	These data also agree well with the previous works, in which the lattice parameters of this particular HEC varies from 4.4 to 4.62 {\AA} depending on the thermal regime of synthesis and chemical composition of raw material \cite{biesuz_interfacial_2020,chicardi_low_2019}.
	There are two unidentified low-intensity reflections at $\sim32^\circ$ and $\sim50^\circ$ which do not correspond to either oxide phases or graphite or other known compounds. 
	
	Thus, the optimal set of parameters for the successful synthesis of a single-phase HEC are the following: mixing time is 300 min, ball-to-powder ratio is 4:1, arc duration is 45 sec, current is 220 A,  temperature $>$ 2000 K. 
	The combination of these parameters ensures homogeneity of the initial mixture of raw material and sufficient temperatures in the entire volume of powder mixture. 
	At other parameters of experiments (longer treatment time, another ball-to-powder ratio, lower current) the product always contains several solid solutions.
	In addition, in the case of formation of multi-phase sample its second treatment with plasma did not lead to the formation of single-phase HEC.
	
	In the low-temperature region of our arc reactor we detected a multi-phase structure.
	As can be seen from the measured XRD pattern the multi-phase sample contains two major phases as evidenced by the presence of additional reflections in the region around 35$^{\circ}$ (see Fig. 3b).
	The lattice parameters of the synthesized phases are equal to $a_1$ = 4.59 $\pm$ 0.02 \AA, $a_2$ = 4.45 $\pm$ 0.01 \AA $ $ (based on a series of more than 30 experiments).
	Low-intensity reflections in the range from 25 to 32$^{\circ}$ correspond to traces of metal oxides. 
	In this case the amount of the injected energy is enough to form an autonomous gas environment and to form the multi-component carbides, but TiZrNbHfTaC$_{5}$ can not be formed at these conditions. 
	
	\begin{figure*}
		\includegraphics[width=0.5\textwidth]{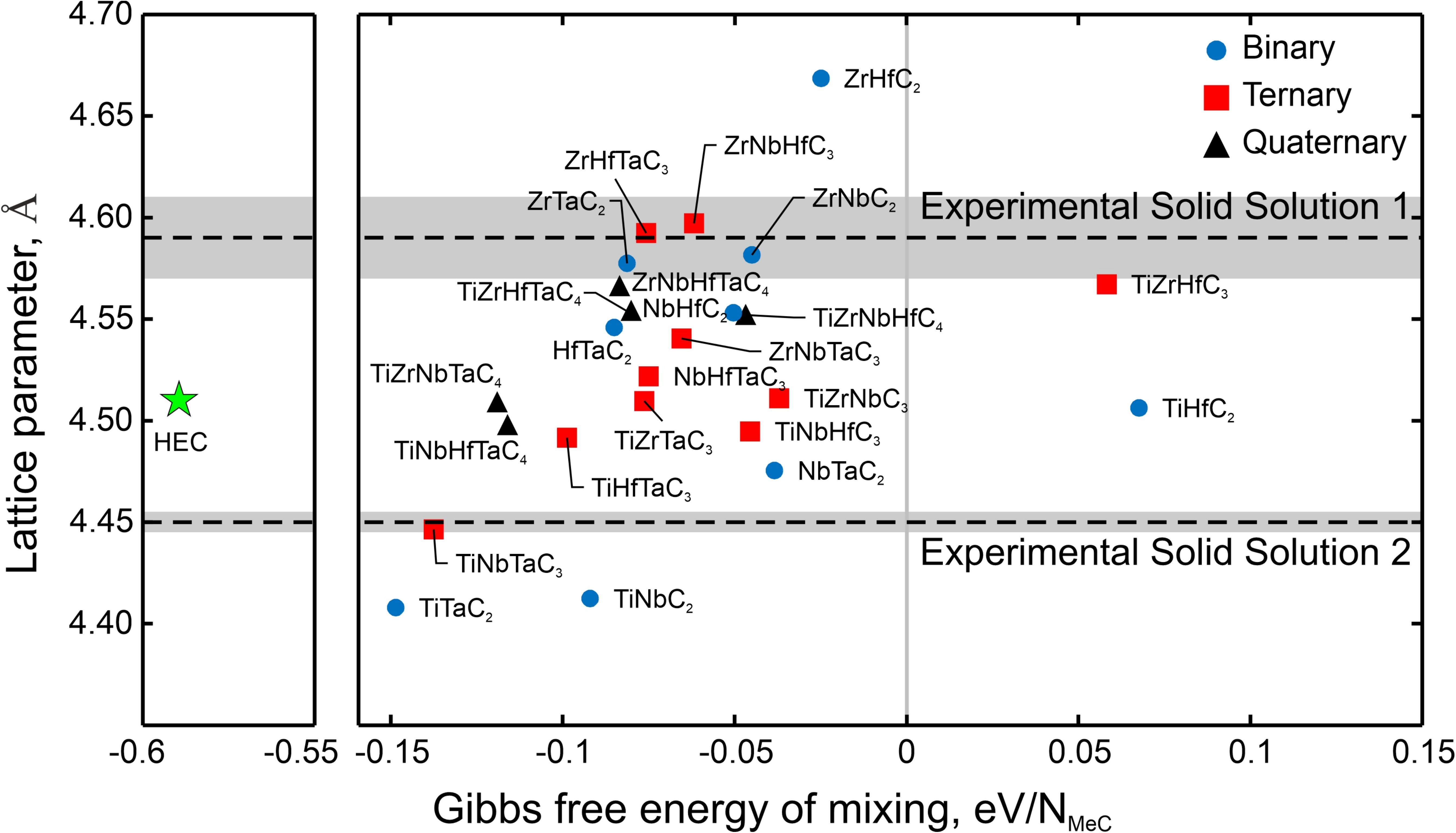}
		\caption{Correlation between the Gibbs free energy of mixing as calculated at 500 K and the lattice parameters of binary (circles), ternary (squares),  quaternary (triangles) carbides and HEC (star). Horizontal dashed lines and shaded areas represent lattice parameters and error of the experimentally observed phases respectively}.
		\label{fig:mix}
	\end{figure*}
	
	To provide a deeper understanding of composition of the phases emerged in a low-temperature region of the sample we performed the analysis of Gibbs free energy of mixing calculated at 500 K. 
	Fig. 4 represents the correlation between Gibbs free energy of mixing of investigated multi-component carbides and their lattice parameter. 
	Information about the calculated Gibbs free energies of mixing at different temperatures for the considered multi-component carbides presented in Table S2 in Supporting Information.
	As it is evident, carbides with the lowest Gibbs free energy of mixing and the best match of lattice parameters with the experimental ones are ZrHfTaC$_{3}$ (experimental solid solution 1) and TiNbTaC$_{3}$ (experimental solid solution 2).
	Their lattice parameters agree well with experimentally determined shown by black dashed lines in Fig. 4.
	However, taking into account the experimental error ($\pm$ 0.02 \AA) in determination of lattice parameters based on 30 measurements we should also consider ZrHfTaC$_{3}$, ZrNbHfC$_3$, ZrNbHfTaC$_4$, ZrNbC$_2$, and ZrTaC$_2$ as potential candidates describing the structure of experimental solid solution 1, see Fig. 4.
	Obtained results of DFT calculations agree well with our assumption based on CMC simulations at 500 K, see Fig. 2a,c.
	The calculated XRD patterns of studied phases are shown in Fig. 3d. 
	Results for TiNbTaC$_{3}$ perfectly matches the experimental XRD pattern of the low-temperature sample verifying that TiNbTaC$_{3}$ is indeed the experimentally observed phases for second solid solution.
	Gibbs free energies of mixing of ZrHfTaC$_{3}$, ZrNbHfC$_3$, ZrNbHfTaC$_4$, ZrNbC$_2$, and ZrTaC$_2$ quite similar within 0.05 eV/N$_{MeC}$. 
	However, calculated lattice parameter for ZrNbHfTaC$_4$ is lower compared to experimental one, which can be clearly seen from XRD pattern in Fig. 3d.
	The rest of considered phases (ZrHfTaC$_{3}$, ZrNbHfC$_3$, ZrNbC$_2$, and ZrTaC$_2$) have very similar lattice parameters with close mixing energies (Fig. 4) resulting in the conclusion about an equal probability of finding any of these phases, or all together, in the experiment as solid solution 1.
	This claim is also supported by the simulated structure, presented in Fig. 2a.
	Evidently, no mixing between Hf/Zr and Ti species was observed leading to the formation of (Ti-Nb-Ta)C and titanium-free (solution of ZrHfTaC$_{3}$, ZrNbHfC$_3$, ZrNbC$_2$, and ZrTaC$_2$ phases) domains.
	
	\subsection{Microscopy analysis of synthesized HECs} 
	
	The detailed study of the two types of samples was performed by transmission and scanning electron microscopy measurements.
	In the Fig. 5 there are two SEM images with energy dispersive X-ray analysis of two samples (single-, and multi-phase). 
	The measured data from the single-phase sample is shown in  Fig. 5a, while the data from the multi-phase sample is shown in Fig. 5b. 
	These data support our results obtained by the XRD analysis (see Fig. 3) in which at least two solid solutions were identified.
	It was observed that the obtained samples contain agglomerates with sizes of 20-60 $\mu$m consisting of individual crystalline particles with sizes of about 1--3 $\mu$m. 
	Individual crystals with larger sizes of 10-20 $\mu$m can be identified as well.
	The identified objects contain carbon, titanium, zirconium, niobium, hafnium, tantalum, oxygen, and minor other impurities. 
	Mapping of chemical elements shows that in the single-phase HEC (Fig. 5a) all chemical elements are almost uniformly distributed in crystal agglomerates and individual crystals.
	In the phase contrast mode it is practically impossible to distinguish several crystalline phases. 
	The obtained images fully confirm the uniform distribution of chemical elements as was shown by our Monte Carlo simulations of the HEC structure at 2000 K (Fig. 2b).  
	
	\begin{figure*}
		\includegraphics[width=0.85\textwidth]{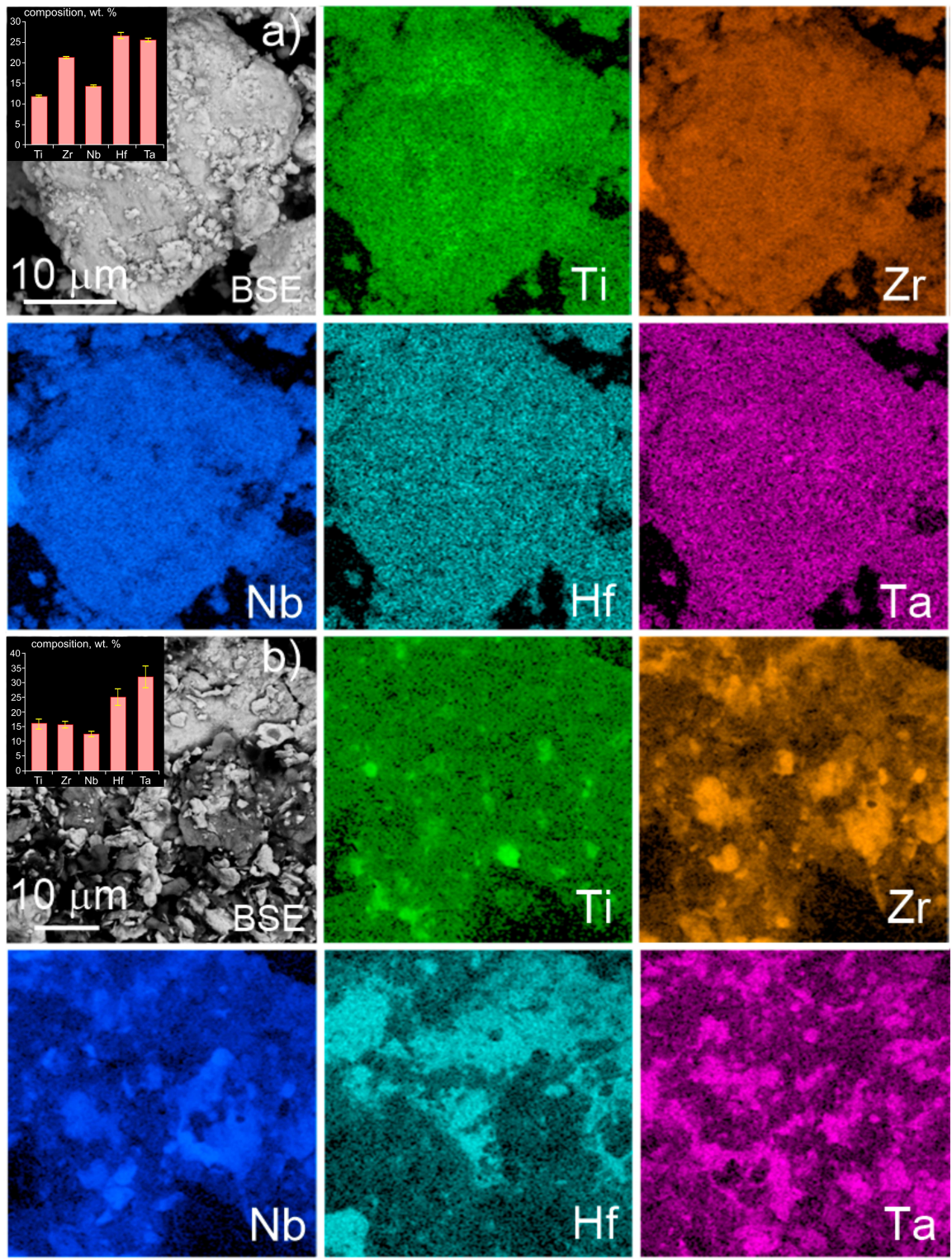}
		\caption{SEM images of experimental samples with energy dispersive X-ray analysis.  a) Single- and b) multi-phase samples. Dispersive X-ray analysis images show the mapping of chemical elements across the sample. 20 points across the sample were used to measure the elemental distributions (shown in the inset)}
		\label{fig:N3}
	\end{figure*}

	The samples with several crystalline phases of metal carbides (Fig. 5b) contain the same chemical elements, but their distribution is irregular, namely the regions with higher densities of individual metals (Ti, Zr, Nb, Hf, Ta) can be observed.
	These images correspond to the simulated structure at 500 K where the phase separation is clearly seen (Fig. 2a).
	Root mean square deviations in the series of measurements of semi-quantitative analysis of chemical composition are significantly higher than in the samples of single-phase HEC (Fig. 5, left bottom inset). 
	
	It should be pointed out that our samples contain $\sim$4 wt.$\%$ of oxygen as it is inevitable for almost any powders. 
	Oxygen comes via the adsorption during the preparation of the raw materials for grinding and during the grinding it interacts with oxygen.
	Moreover, presence of some oxygen is a general characteristic of the HEC samples obtained by using other methods, as was noted above.
	We assume that majority of the oxygen is presented in the surfaces of synthesized particles.
	Anyway presence of a few percent of oxygen does not contradict the conclusions about the synthesis of HEC.

	Further detailed investigation of crystal structure of two samples was made by scanning and transmission electron microscopy (STEM) as shown in Fig. 6 and Fig. 7.
	As can be seen from obtained images of the single-phase sample (Fig. 6), there are agglomerates with the average size of 0.5-1.0 $\mu$m.
	Each agglomerate consists of the nanoparticles with the sizes from tens to hundreds of nanometers.
	Electron diffraction image clearly shows the single-phase system with measured lattice spacing of 2.695 ${\pm}$ 0.056 \AA, (111) plane, 1.676 ${\pm}$ 0.014 \AA, (220) plane, 1.399 ${\pm}$ 0.035 \AA, (222) plane, and 1.072 ${\pm}$ 0.027 \AA, (210) plane. 
	These lattice spacing corresponds to lattice parameter of studied single-phase HEC, $a$ = 4.49 $\pm$ 0.03 \AA.
	The obtained data agrees well with the calculated ones for the single-phase HEC as presented in Table S3 (see Supporting Information).
	
	\begin{figure*}
		\includegraphics[width=0.5\textwidth]{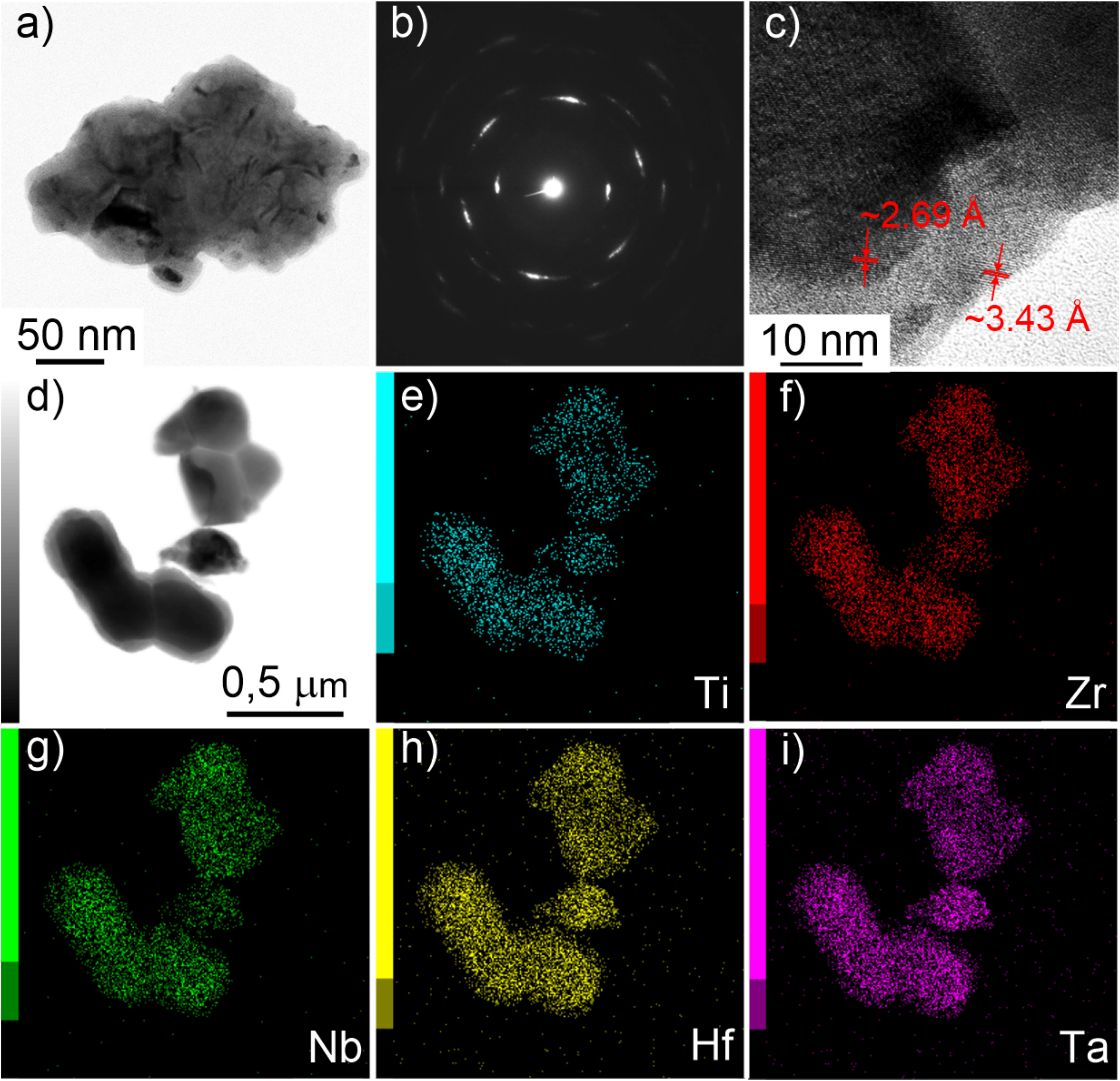}
		\caption{Microscopy images of a single-phase sample. a) TEM image, b) SAED, c) HRTEM lattice image, d) STEM, e-i) EDX elemental mapping.}
		\label{fig:N4}
	\end{figure*}
	
	\begin{figure*}
		\includegraphics[width=0.5\textwidth]{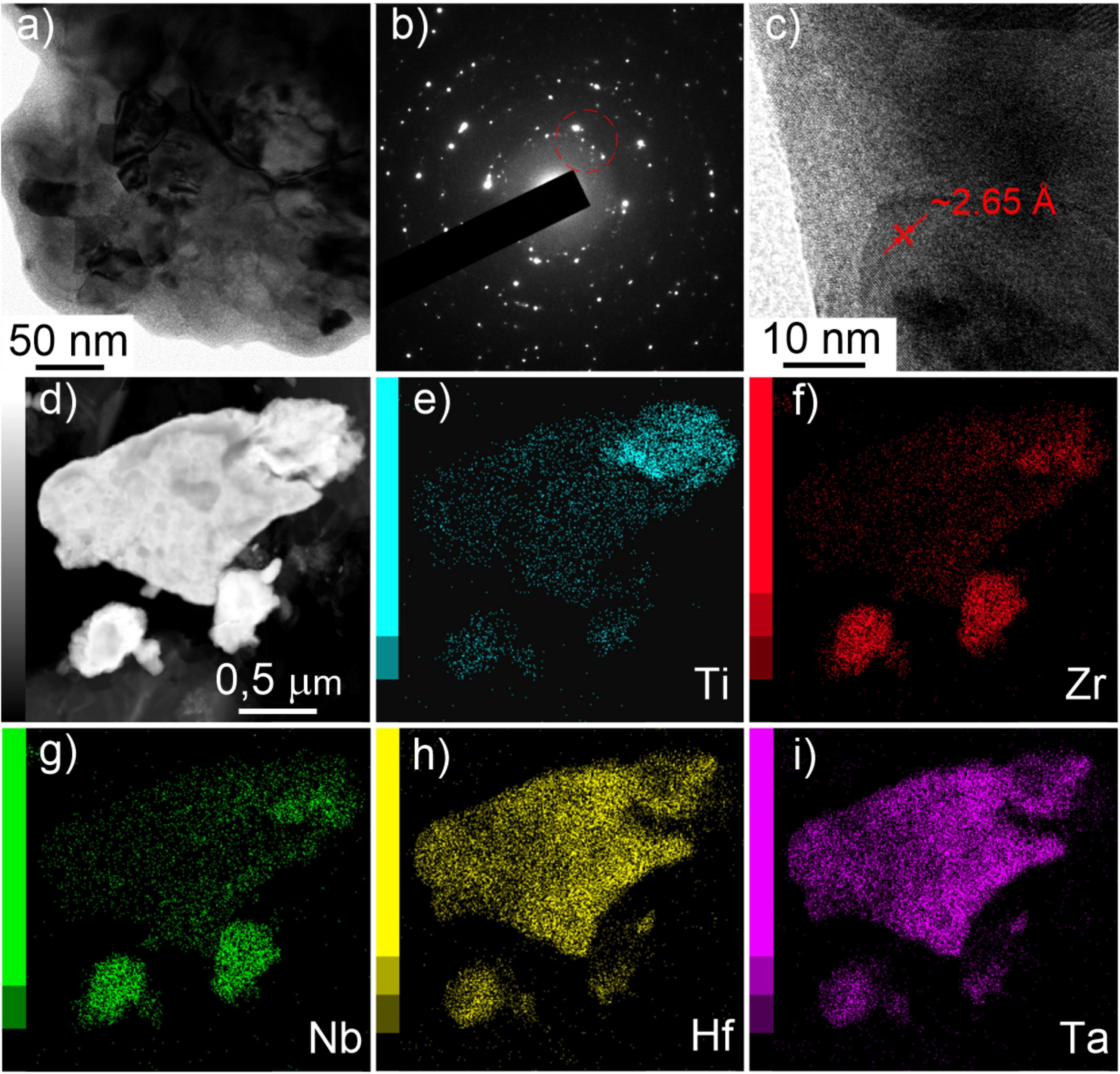}
		\caption{Microscopy images of a multi-phase sample. a) TEM image, b) SAED, c) HRTEM lattice image, d) STEM, e--i) EDX elemental mapping.}
		\label{fig:N5}
	\end{figure*}

	According to the obtained lattice images one can observe the \enquote{core-shell} structure in the synthesized HEC particles.
	There is interplanar spacing of 2.69 {\AA} observed in the \enquote{core}, while in the \enquote{shell} we observed the interplanar spacing of 3.43 {\AA}.
	Electron diffraction pattern measured from the \enquote{core} of observed particle shows the presence of reflections related to the interplanar spacing of  2.11, 1.63, and 1.40 {\AA}.
	Thus it can be concluded that the \enquote{core} of the particle is made of TiZrNbHfTaC$_{5}$, while \enquote{shell} is a graphite.
	This can be also found from the EDX elemental mapping (Fig. 6e-i), where the uniform distribution of elements was observed.
	Such a structure can be formed due to differences in the melting temperatures of carbon and transition metals carbides. 
	Graphite surrounding the carbide particles is a common situation for the arc plasma synthesis and was observed previously in Refs.\ \cite{saito_encapsulation_1993,saito_encapsulation_1997}.
	The elemental mapping in Fig. 6 shows a uniform distribution of metals (Ti, Zr, Nb, Hf, Ta) in the formed agglomerates, which also confirm the presence of the single-phase HEC.
	
	In the samples containing at least two carbide phases (according to the XRD data), Fig. 7, the agglomerates with the size of $\sim$150 nm are covered by a carbon shell.
	The separate grains with the size of $\sim$10 nm can be observed inside each agglomerate.
	According to the SAED measurements one can clearly see the presence of several crystalline phases with close lattice parameters.
	In particular, reflections from the area of the first diffraction ring correspond to the interplanar spacing from 2.66 to 2.72 {\AA}.
	One can also identify the averaged interplanar spacing of 2.31, 1.88, 1.64, 1.40, 1.33, 1.25, 1.15, and 1.04 {\AA}, the population of which may indicate the presence of several cubic carbide phases in the samples (Fig. 7b).

	From the HRTEM lattice images we can identify interplanar spacing  of about 2.65 {\AA}, which may correspond to the family of (111) planes of the one of the solid solutions.
	According to the elemental mapping (Fig. 7e--i) we can see that the chemical elements are distributed unevenly, with individual metals dominating in individual grains of sample indicating the presence of several phases of solid solutions of cubic phases. 
	Thus, the results of the transmission electron microscopy confirm the conclusion from X-ray diffraction and scanning electron microscopy about presence of several carbide phases in these multi-phase sample. 
	
	\subsection{Oxidation resistance} 
	Transition metal carbides usually display high melting temperatures together with high oxidation resistance. Synthesized samples should also possess pronounced stability with respect to oxidation. 
	To study this, we have performed the thermogravimetric analysis (TGA), differential analysis (DTG), and differential scanning calorimetry (DSC) in the air.
	The obtained results are shown in Fig. 8. 
	According to the obtained  TG-curves (Fig. 8a) two exothermic processes can be identified; first process is accompanied by mass gain, and the second one is by mass loss. 
	The DTG analysis (Fig. 8b) shows that the highest rate of mass gain for a single-phase sample (red dashed line) corresponds to the temperature of 620$^{\circ}$C, which is higher than the corresponding value of multi-phase sample (570$^{\circ}$C, blue solid line).
	Multi-phase sample begins to oxidize at 496$^{\circ}$C, while oxidation of single-phase sample occurs at higher temperature of 535$^{\circ}$C.
	The highest oxidation rate with mass gain equal to 1.13 wt.\%/min corresponds to multi-phase sample at 570$^{\circ}$C.
	The highest Oxidation rate of single-phase sample at 625$^{\circ}$C is 1.11 wt.\%/min.
	Thus, for the single-phase sample, the temperature corresponding to the maximum oxidation rate is 55$^{\circ}$C higher compared to multi-phase sample.
	According to the DSC analysis (Fig.\ 8c) both processes of mass gain and mass loss occur with the release of energy.
	The DSC curve of single-phase sample is characterized by the maximum at temperature about 620$^{\circ}$C, while the second one is weakly intensive, at the temperature 780$^{\circ}$C .
	The DSC curve of the multi-phase sample is characterized by several maxima at 580, 700, 820, 900$^{\circ}$C, while the most intensive one corresponds to the temperature 580$^{\circ}$C.  
	Similar results were obtained in Ref.\cite{zhou_high-entropy_2018} where the oxidation behavior of multi-phase samples of binary metal carbides and single-phase HEC TiZrNbHfTaC$_5$ obtained from the binary ones.
	It was found that the oxidation of single-phase HEC occurs with a characteristic monomodal DSC peak at higher temperature with respect to the multi-phase powder.
	In general, oxidation behavior of HECs is an understudied issue to date.
	Nevertheless, this trend, according to the literature, is also typical for HEC of other compositions: HEC ZrNbHfTaC$_4$ is characterized by a higher oxidation temperature and a temperature corresponding to the maximum oxidation rate compared to the same characteristics of binary metal carbides \cite{wang_oxidation_2021}
	
	\begin{figure*}
		\includegraphics[width=\textwidth]{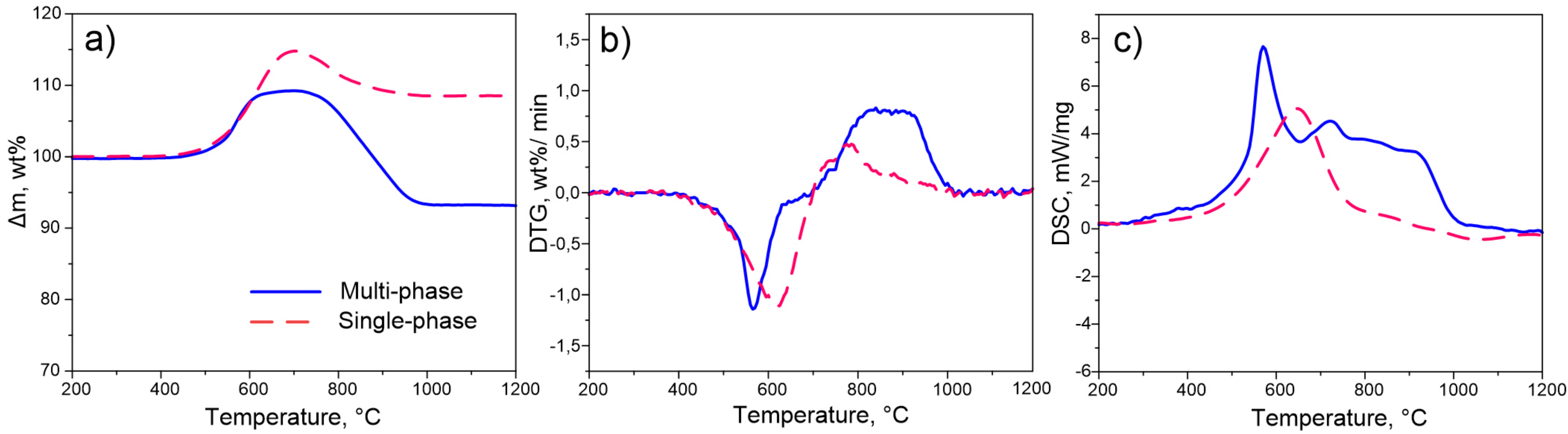}
		\caption{Results of differential thermal analysis. a) TG, b) DTG, and c) DSC in the air for two types of samples, namely multi-phase (blue solid curves) and single-phase (red dashed curves).}
		\label{fig:N6}
	\end{figure*}
	
	In conclusion, according to the differential thermal analysis, the first exothermic effect with mass gain can be identified as a combustion reaction of carbide phases.
	It was found that the temperature of the highest intensity of combustion of carbide phases is higher in the single-phase sample.
	The DSC data show that the single-phase sample corresponds to the monomodal maximum of exothermic process, while the DSC data of the multi-phase sample clearly shows several exothermic processes.
	The DSC data indirectly confirm the presence of several crystalline phases in the multi-phase sample and one in the single-phase samples. 
	It is assumed that the single-phase HEC should be characterized by a higher oxidation temperature than a mixture of other carbides consisting of the same chemical elements \cite{wang_oxidation_2021,wang_oxidation_2021_2}. 
	Thus, the data of the differential thermal analysis performed in the air is consistent with the data of other analytical techniques and literature data. 
	It should be noted that the oxidation products of HEC of considered composition have been previously studied in Ref.\cite{zhou_high-entropy_2018} at different temperatures to be metal oxides.
	The developed method, based on a combination of theoretical and experimental techniques, allowed to adjust electric arc plasma parameters and perform a targeted synthesis of single-phase TiZrNbHfTaC$_5$ carbide within theoretically defined temperature regimes. 
	
	Indeed, the implications and importance of this work extend far beyond the results shown herein.
	There are over 60 currently known high-entropy carbides and thus this work opens the door for the development of a large number of various materials, including the carbides, nitrides, borides of transition metals.

	\section{Methods}
	\subsection{Precursor selection and characterization}
	Commercially available metal powders (Ti, Zr, Nb, Hf, Ta) with purity not less than 99.9$\%$ and avarage size $\leq$ 10 $\mu$m (Rare Metals corp., Russia) were used as a raw material.
	An ultradispersed carbon powder with a purity no worse than 99.9$\%$ with average size $\leq$ 1 $\mu$m (Hi-tech Carbon Co, China) was used as a source of carbon.
	Powders were analyzed using X-ray diffraction analysis by Shimadzu XRD 7000s, with wavelength of 1.54060 \AA, with built-in Shimadzu software (the results are given in Supplementary Fig. 3). 
	
	\subsection{Ball milling}
	Equimolar mixtures of initial powders were prepared and mixed in a ball mill (Mill 8000M Horiba Scientific).
	The ball milling procedure was designed to mix the powders of raw materials in order to homogenize the powder mixture.
	Equipment and balls were made of zirconium dioxide and a ball-to-powder ratio was 4:1.
	Separate experiments were performed with preliminary grinding of the initial raw material with balls of tungsten carbide with  a ball-to-powder ratio of 2.5:1.
	Several series of experiments were performed with mixing time from 45 to 540 min.
	Each obtained equimolar mixture was treated in different experiments with arc durations of 15, 30, 45, and 60 sec, at different currents of 50, 100, 150, and 200 A.
	
	\subsection{Arc plasma synthesis}
	The synthesis strategy of high-entropy carbides consists of applying electric arc plasma to raw material containing metals and carbon in an autonomous gas environment.
	In this process oxygen binds to carbon to form CO and CO$_{2}$ gases, which form an autonomous gas environment.
	According to special design of the discharge circuit electrodes in conjunction with the operating modes of the plasma reactor it is possible to realize the following reactions at arc discharge in air medium:
	\begin{equation}
		\begin{split}
			& Ti+C \rightarrow TiC  \\ 
			& Zr+C \rightarrow ZrC     \\
			& Nb+C \rightarrow NbC   \\
			& Hf+C \rightarrow HfC   \\
			& Ta+C \rightarrow TaC    \\
			& 2C + O_{2} \leftrightarrow 2CO   \\ 
			TiC + ZrC + NbC + & HfC  + TaC \leftrightarrow TiZrNbHfTaC_{5}   \\ 
		\end{split}
	\end{equation}
	
	To perform synthesis of single- and multi-phase high-entropy carbide we used the arc plasma reactor which simplified schematic illustration is shown in Supplementary  Fig. 8a. 
	The anode is represented as a graphite rod (diameter is 8 mm, length is 100 mm), the cathode is a graphite crucible with diameter of 25 mm (images of these dishes are shown in Supplementary Fig. 9).
	We use commercially available graphite electrodes with purity not less than 99.9$\%$ (QiJing Trading Co, China).
	The initial raw material is loaded into the cavity of the graphite crucible and covered with a lid. 
	The anode was lowered into the cavity of the graphite crucible by using a linear electric drive, in which an arc discharge was activated and maintained (arc gap was about 1 mm). 
	Rectifier/inverter converter with operating current from 20 to 220 A was used as a power source.
	The temperature of the raw material was recorded using W-Re thermocouples, the signal from which was transmitted to a specialized controller.
	The composition of the forming gas medium was studied earlier for similar reactors and their operating modes\cite{pak_vacuumless_2020}.  
	
	It is known that the thermal field of the DC arc plasma is characterized by a high temperature gradient. Locally in the arc binding zone the temperature can reach 10 000$^{\circ}$C \cite{wang_numerical_2020,zhong_deep_2020}, and at a distance of 8--10 mm across the electrode surface the temperature decreases to 2000--3000$^{\circ}$C (at the current of 200 A) \cite{wang_numerical_2020,zhong_deep_2020}.

	Synthesis of metal carbides and HECs usually requires temperatures of about 1900--2500$^{\circ}$C \cite{zhou_electromagnetic_2021,wei_high-entropy_2021,wang_oxidation_2021}. 
	The results of temperature measurements in realized conditions showed that at the current of 50 A during the working cycle of 45 seconds the temperature at a given point reaches $\sim$1000$^{\circ}$C, at the current of 100 A the temperature reaches $\sim$1250$^{\circ}$C, at the current of 150 A the temperature reaches $\sim$2100$^{\circ}$C, and at 200 A the maximum registered temperature is 2150$^{\circ}$C. 
	The dependence of achieved maximum temperature on the time is shown in Supplementary  Fig. 8b.
	At higher temperatures the thermocouples destroyed. 
	This happens because corundum cover of the thermocouple is damaged and there is also a damage in the registration circuit, which prevents the measurement of higher temperature values in the system. 
	Plasma duration of 45 second at the current of 220 A is critical and close to the maximum allowable, because a longer plasma duration causes overheating of current-carrying parts in the area of contact with the graphite electrodes, and they melt.
	Based on obtained data and our actual experimental measurements, we conclude that our design of arc reactor provides the necessary conditions for the formation of high-entropy metal carbides.  
	
	\subsection{Morphology, homogeneity, and purity characterization}
	Qualitative X-ray diffraction analysis of synthesized samples was performed by using Shimadzu XRD 7000s, with wavelength of $\lambda$ = 1.54060 \AA, with built-in Shimadzu software. 
	The morphology of the micro-sized objects in the products was examined by using a Tescan Vega 3 SBU scanning electron microscope equipped with an Oxford X-Max 50 energy dispersive X-ray analysis (EDX) attachment with a Si/Li crystal detector.
	Transmission electron microscopy was performed on a JEOL JEM 2100F transmission electron microscope with an energy dispersive analysis attachment.
	Samples were prepared in an alcohol suspension in an ultrasonic bath, applied to a standard copper grid coated with an amorphous carbon.
	
	\subsection{The low-rank potential model}
	For modeling interatomic interactions we employed the low-rank potential (LRP) model \cite{SHAPEEV201726}, which is a machine-learning interatomic potential, trained on a set of quantum-mechanical data.
	In the LRP model atomistic structure is represented by an ideal crystal lattice with sites, occupied by one of the chosen atomic species (Ti, Zr, Nb, Hf, and Ta in our case).
	The carbon atoms are modeled with LRP implicitly: they are not degrees of freedom as they are always present at the 4b sites, however their interaction is present in the energies that the LRP is trained on.
	The set of occupied neighbors closest to the chosen site is called the neighborhood of this site (Supplementary Fig. 10).
	Hence, the neighborhood of each site has a contribution to the total energy of the atomic configuration given by the formula:
	\begin{eqnarray}\label{eq:KL}
		V(\xi) = V(\sigma_{1},...,\sigma_{n}) = V(\sigma(\xi + r_{1}),...,\sigma(\xi + r_{n})),
	\end{eqnarray}
	where $V$ is the LRP model in the tensor form, $\xi$ is the position of the central atom, $\sigma(\xi + r_{i})$ is the atomic type of \textit{i}th site; \textbf{r} is the vector, connecting the central atom with the $i$th neighbor, and $n$ is the number of closest neighbors (see Supplementary Fig. 10).
	Subsequently, the total energy of the atomic configuration is written as:
	\begin{eqnarray}\label{eq:KL}
		E = \sum_{\xi \in \Omega} V(\sigma(\xi + r_{1}),...,\sigma(\xi + r_{n})),
	\end{eqnarray}
	where $\Omega$ denotes periodically repeated in space lattice sites . 
	
	The tensor $V$ contains energy contributions of all possible, $m^{n}$, atomic neighborhoods, where $m$ is the number of atomic species.
	Such tensor would consist of more than one billion parameters which is completely unfeasible to obtain from quantum-mechanical calculations.
	In order to reduce the number of parameters, the tensor-train decomposition is applied \cite{ttrain}.
	Within this formalism, we assume that:
	\begin{eqnarray}\label{eq:KL}
		V(\sigma_{1},...,\sigma_{n}) = \prod_{i}A_{i}(\sigma_{i}),
	\end{eqnarray}
	where $A_{i}$ are matrices with rank $r$ or less, which contain the parameters of the LRP model. The size of $A_{1}$ and $A_{n}$ is $1\times r$ and $r\times1$ respectively, and the sizes of $A_{2},...,A_{n}$ are $r\times r$, so that their product gives a scalar, which corresponds to the energy contribution of a neighborhood, labeled by atomic species $\sigma_{1},...,\sigma_{n}$. Eventually, such decomposition allows us to reduce the number of parameters from $m^n$ to $nmr^2$. 
	\\
	Thus, the predictive accuracy of the LRP model can be controlled by two adjustable parameters, namely, rank $r$ and the number of neighbors $n$.
	Here we restricted the interaction to the nearest neighbors only, $n=13$, as $fcc$ lattice has coordination number (CN) of 12, and $n = CN + 1$ (including the central atom of the neighborhood).
	The value of rank $r$, was set to $r=3$, which eventually gave us about 600 independent parameters in the LRP model.
	The parameters can be found by solving the minimization problem with the following functional:
	\begin{eqnarray}\label{eq:KL}
		\frac{1}{K}\sum_{k=1}^{K}|E(\sigma^{(k)}) - E^{qm}(\sigma^{(k)})|,
	\end{eqnarray}
	where $\sigma^{(k)}$ are the atomic configurations, with the total number of $K$ in the training set, and $E(\sigma^{(k)})$ and $E^{qm}(\sigma^{(k)})$ are the energies of $\sigma^{(k)}$ calculated by LRP and DFT, respectively. 
	The minimization is done by the alternating least (ALS) squares method, which in our case simply optimizes one matrix $A_{i}$ at a time, and simulated annealing that adds random Gaussian noise to every element of $A_{i}$, which decreases from one ALS iteration to the next one.
	
	The workflow for the LRP training is presented in Fig.\ S1a. 
	As it is seen from the algorithm, the initial dataset consisted of 150 random configurations, each of which was a $2\times 2 \times 2$ supercell (32 atoms) and they were confined to be as equimolar as possible.
	After that, we performed canonic Mote Carlo simulation from which we sampled additional configurations and added them to the initial dataset.
	During each training stage, we split the dataset into training and validation subsets of sizes 80\% and 20\% of the initial set respectively.
	After performing several iterations of the algorithm depicted in Supplementary Fig. 1a, 250 new configurations were added to the initial dataset.
	Eventually, this allowed us to achieve validation error of 9 meV/atom of the LRP, that we consider sufficient.

	\subsection{Monte Carlo Simulation}
	We have performed canonic Monte Carlo (CMC) simulation \cite{cmc} coupled with LRP for modeling the thermodynamic equilibrium structures at different temperatures.
	Specifically, we focus on temperature range from 500 to 2000 K, which is relevant to the experiment.
	Simulation is carried out for quite small initial configurations with 865 metallic atoms ($10\times 12\times 15 $ supercell), which randomly occupy 4\textit{a} Wyckoff positions of the face-centered cubic lattice. 
	During each CMC step the new structure is generated by an interchange of metallic atoms of different chemical types on randomly chosen nearest-neighbor sites of the lattice (Supplementary Fig. 1b).
	The new structure is accepted if its energy is lower in comparison with previous one.
	Otherwise, the algorithm generates random number $\gamma \in$ [0, 1] and accepts the new structure under the following condition:
	\begin{eqnarray}\label{eq:mean_E}
		\gamma < \exp{(\frac{E_{i-1} - E_{i}}{kT})},
	\end{eqnarray}
	where $E_{i-1}$ and $E_{i}$ are the energies of the previous and the current structures.
	
	Convergence of the simulation is evaluated based on the mean energy value, given by:
	\begin{eqnarray}\label{eq:mean_E}
		\bar{E} = \frac{1}{n}\sum_{i=1}^{n}E_{i},
	\end{eqnarray}
	where $n$ - number of annealing steps. The simulation is converged when the change of $\bar{E}$ between two subsequent steps is below $10^{-3}$ eV. 
	On practice we set the number of steps to $2 \times 10^8$ for each temperature, which is sufficient enough to meet the convergence criterion.
	
	\subsection{Specific heat capacity}
	In our calculations we consider only configurational part of specific heat capacity. In particular, it measures the structural changes that occur at a certain temperature. Specific heat capacity is calculated as:
	\begin{eqnarray}\label{eq:spec_heat}
		C_\text{V}(T) = \frac{\sigma^{2}}{n(kT)^{2}},
	\end{eqnarray}
	where $\sigma^{2}$ - energy variance, calculated over a number of CMC steps; $n$ - number of atoms; $k$ - Boltzmann constant; T - heating temperature.
	
	\subsection{Miscibility analysis}
	To determine the phases, that might occur during sample decomposition the miscibility of binary, ternary, and quaternary high-entropy carbides from the mixture of individual metal carbides (TiC, ZrC, NbC, HfC, TaC) was analyzed by the Gibbs free energy of mixing $G_{mix} = H_{mix} - TS_{mix}$ \cite{liu_phase_2021, liu_phase_2022}. 
	Here $H_{mix}$ is the enthalpy of mixing that can be calculated as follows:
	\begin{eqnarray}\label{eq:Emix}
		H_\text{mix} = \left( E_\text{tot} - \sum_i N_i E_i^{\textrm{MeC}} \right)  / \sum_i N_i,
	\end{eqnarray}
	where $E_{tot}$ is the total energy of the considered multi-component carbide, $E_i^{\textrm{MeC}}$ is total energy of individual metal carbide, $N_i$ is the number of individual metal carbides.
	
	Mixing entropy $S_{mix}$ in the homogeneous limit can be determined by using the Boltzmann's entropy formula
	
	\begin{eqnarray}\label{eq:Smix}
		S_\text{mix} = k_\text{B} \sum_{i=1}^N x_i \ln x_i,
	\end{eqnarray}
	where $k_B$ is the Boltzmann constant, $x_i$ is the mixing concentration (0.5 for binary carbides, 0.33 for ternary carbides, and 0.25 for quaternary carbides), $N$ is a number of metal species in considered carbides (2, 3, 4 for binary, ternary and quaternary carbides respectively).  
	
	To perform such calculations we generated 2$\times$2$\times$2 supercells with the rocksalt crystal structure having 32 metal atoms and 32 carbons. 
	Generation of disordered distribution of metal atoms across the supercell was made by using special quasirandom search algorithm \cite{oganov_how_2009, lyakhov_crystal_2010}.
	Maximising disorder in this situation, one gets a generalized version of the special quasirandom structure.
	Thus, there are 5 quaternary, 10 ternary, and 10 binary carbide structures were generated. 
	All generated structures together with individual carbides (TiC, ZrC, NbC, HfC, TaC) were relaxed with VASP \cite{VASP1, VASP2, VASP3}.
	
	\subsection{Density functional theory calculations}
	The LRP \cite{SHAPEEV201726}, which was used as an interaction model in Monte Carlo method, was trained on DFT calculations.
	To compute reference energies for the LRP training, VASP 5.4.4 \cite{VASP1, VASP2, VASP3} was used.
	In our calculations the projector augmented wave (PAW) \cite{paw} method utilizing the Perdew-Burke-Ernzerhof generalized gradient approximation (PBE-GGA) \cite{gga} was employed.
	For the training set we generated $2 \times 2 \times 2$ (32 atoms) supercells, based on rocksalt $fcc$ unit cell.
	The value of plane-wave cutoff energy was set to 540 eV, which is 1.9 times larger than the highest ENMAX energy of the utilized PAW pseudopotentials.
	We generate $4 \times 4 \times 4$ \textit{k}-point mesh using the Monkhorst-Pack scheme \cite{mon-pack}.
	To account for the impact of lattice relaxations, both ionic and cell relaxations were included. 
	The energy convergence criteria for these types of relaxations was set to $10^{-5}$ eV.
	
	To take into account vibrational contribution at different temperatures to the mixing energies of considered individual, binary, ternary, quaternary carbides along with TiZrNbHfTaC$_{5}$ we have calculated the Gibbs free energy of mixing instead of mixing enthalpy $H_{mix}$ (eq. \ref{eq:Emix}) as follows:
	\begin{eqnarray}\label{eq:Gmix}
		G_\text{mix}\left( T \right) = \frac{ G_\text{tot} \left( T \right) - \sum_i N_i G_i^{\textrm{MeC}}\left( T \right)} {\sum_i N_i},
	\end{eqnarray}
	The Gibbs free energy of considered structure ($G_\text{tot}$) and individual carbide ($G_i^{\textrm{MeC}}$) can be calculated by using the following formula:  
	\begin{eqnarray}\label{eq:G}
		G\left( T, V \right) = E_0 \left(V \right) + F_\text{vib} \left( T, V \right) - TS_\text{conf} ,
	\end{eqnarray}
	where $E_0$ is the total energy from the DFT calculations, $TS_\text{conf}$ is the configurational entropy, $F_\text{vib}$ is the vibrational Helmholtz free energy calculated from the phonon density of states by using following relation in the harmonic approximation\cite{kern_ab_1999}:
	\begin{equation}\label{eq:Helm}
		\begin{split}
			F_\text{vib} \left( T, V \right) = k_\text{B} T  \int_{\Omega}^{} g \left( \omega \left( V \right) \right) \times \\ 
			\times \ln{\left(1 - \exp{- \frac{\hbar \omega \left( V \right)}{k_B T}}\right)}\,dx + \\
			+ \frac{1}{2} \int g \left( \omega \left( V \right) \right) \hbar \omega \,d\omega,
		\end{split}
	\end{equation}
	Here $ g\left( \omega \left( V \right) \right)$ is the phonon density of states at a given volume, calculated using the finite displacements method as implemented in PHONOPY \cite{togo_first_2015,togo_first-principles_2008} with forces computed using VASP \cite{VASP1, VASP2, VASP3}.
	
	\section{Data Availability Statement}
	Minimal dataset required for the reproduction of results are available via link mentioned in software policy checklist. Access is granted upon request.
	
	\section{Code Availability Statement}
	The software required for the reproduction of results are available via link mentioned in software policy checklist. Access is granted upon request.
	
	\section{Acknowledgements}
	A.G.K. acknowledges financial support from the Russian Science Foundation (grant No 19-72-30043) for performing first-principles calculations of studied compounds. V. S. and A.V.S acknowledge financial support from the Russian Science Foundation (grant No 18-13-00479) for developing the code for CMC and LRP and performing simulations for TiZrNbHfTaC5.
	A.Y.P and Y.Z.V acknowledge support from the Russian Science Foundation (grant No 21-79-10030) for performing the vacuumless synthesis of high-entropy carbides and examination of their properties.
	Authors thank the Tomsk Polytechnic University development program (DC arc plasma reactor system’s automation).
	
	\section{Author Contributions}
	$^{\dag}$ This authors contributed equally to this work.
	V.S. and A.V.S. developed the code for CMC with LRP and performed the simulations of TiZrNbHfTaC5.
	A.Y.P., A.A.G., Y.Z.V., and G.Y.M. performed experimental synthesis of HEC samples and microscopy analysis.
	Z.S.B. performed thermal analysis o f samples.
	A.G.K. and Y.A.K. prepared the theoretical analysis and miscibility analysis.
	V.S., A.Y.P, and A.G.K. wrote the first draft of the manuscript.
	All the authors provided critical feedback and helped shape the research.
	
	\section{Competing Interests}
	The Authors declare no Competing Financial or Non-Financial Interests.

	\bibliographystyle{naturemag}
	\bibliography{main}
\end{document}